\documentclass{osa-article}

\journal{ao}

\articletype{Research Article}

\begin{document}

\title{Smartphone-based measurements of the optical properties of snow}

\author{Markus Allgaier,\authormark{1,*} and Brian J. Smith\authormark{1}}

\address{\authormark{1}Department of Physics and Oregon Center for Optical, Molecular, and Quantum Science, University of Oregon, Eugene, Oregon 97403, USA}

\email{\authormark{*}markusa@uoregon.edu}

\begin{abstract}
Snow is a highly complex medium composed of ice crystals of various shapes and sizes. Knowledge of its intrinsic optical properties such as the scattering and absorption coefficient is tantamount to radiative transfer models in climate research. The absorption coefficient, in particular, allows us to access information about light-absorbing particles contained in the snow. In contrast to snow's apparent properties like the albedo, measuring the intrinsic properties is challenging. Here, we present a simple apparatus that can measure bulk optical properties of snow using readily available components and a smartphone camera, and a robust diffuse-optical framework for data analysis. We demonstrate the instrument both on scattering phantoms with known scattering and absorption coefficient as well as in the field. Its low cost, simplicity and portability uniquely qualify this setup for large-scale field work, undergraduate education and citizen science.
\end{abstract}

\section{Introduction}
Snow covers large part of the planet in winter and plays a fundamental role in the water cycle. Especially in regions experiencing dry summers, snow acts as a frozen reservoir that provides a constant supply of melt water \cite{mote2018}. Predicting snow melt relies on exact knowledge of the absorption of solar radiation, which is governed by its albedo \cite{nolin1997,skiles2017,zeitz2021}. Empirical albedo models are powerful predictors for melt models \cite{dalum2020,pedersen2005}. However, the intrinsic optical properties of snow are rarely known, and the few data sets that exist focus on the absorption coefficient of ice \cite{wiscombe1980}. Recently, the field has received more attention \cite{wuttke2006,michalsky2013,picard2020}. The absorption coefficient of snow and therefore its albedo depend on light-absorbing particles (LAPs) in the snow \cite{warren1980b,hadley2012,saito2019,beres2020}. The optical absorption coefficient links the apparent properties such as albedo and the chemical composition (LAP content) so that knowledge of the absorption coefficient allows to infer both. Therefore, measuring it would present an opportunity to obtain effective LAP concentrations in the field in order to study the spatial distribution of LAPs \cite{doherty2014,rowe2019,zhong2019}. This question is of particular interest in the light of recent extreme wildfire activity all over the world, which leads to increased LAP concentration and albedo reduction in the snow \cite{gleason2013,kaspari2015}, an effect which lasts many years \cite{gleason2019}.
Likewise, the scattering coefficient presents the link between the physical structure expressed in terms of grain size, grain shape and density\cite{skiles2017}, with the albedo.

Unlike the microscopic properties (grain shape and size, chemical composition), intrinsic bulk properties (absorption and scattering coefficient) allow for a palatable description of the albedo \cite{kokhanovsky2004} without knowledge of the particular scattering behavior of the individual snow grain. For this description, excellent agreement with field measurements of the albedo were found \cite{kokhanovsky2018,picard2020}. To date, the only feasible method of measuring bulk absorption and scattering coefficients is through transmission spectroscopy in shallow boreholes \cite{warren2006}, which requires extensive equipment and stable lighting conditions. When it is compared to that of clean ice, the ice absorption coefficient expresses the absorptive contribution to albedo reduction, independent of what impurities are responsible. For context this increase in absorption can be attributed to any impurity such as black carbon (BC) in order to provide some additional context or interpretation, even though it is unlikely that the impurity content is entirely composed of BC. In comparison, soot photometers are capable of measuring the true black carbon concentration \cite{schwarz2012}, a metric that fails to encompass the effect of all impurities on albedo reduction, which is ultimately the metric of interest.

Here, we present a smartphone-based sensor that can measure the optical scattering and absorption coefficient in the field. The setup is cost-efficient and highly portable. It uses a differential measurement at two wavelengths based onto diffuse propagation of light from a point source in a scattering medium \cite{allgaier2021}. One wavelength (650 nm) serves as a reference point at the red end of the visible spectrum, where absorption is dominated by ice rather than LAPs \cite{warren2008}. Using this reference allows us to extract the absorption coefficient of the ice itself as well as the snow as a bulk material at any visible wavelength. The setup requires only two low-cost diode laser modules and a camera. We use a smartphone camera for simplicity, a detector available to most and previously exploited to fashion colorimeters, photometers and spectrometers \cite{nonno2021}. After using laboratory scattering samples for verification of the method, we deploy the sensor in the field in the Central Oregon Cascades to measure the scattering and absorption coefficients at 405 nm, where the absorption of ice is close to its minimum. Optical field measurements are supplemented by visual estimation of snow grain size and shape as well as density measurements. We calculate the spectral albedo at two wavelengths, 405 nm and 650 nm, and compare it to the spectral albedo measured directly with a spectrometer as well as that simulated using the inferred effective impurity calculation.

This smartphone-based sensor can offer data that was previously hard to obtain, that is of high quality and that can be used to infer both the apparent and microscopic properties of snow. This will enable researchers to conduct extensive field work without relying on chemical analysis or expensive equipment.

\section{Optical bulk properties of snow}

Radiative transfer in scattering media is governed by properties of both the single particles as well as the bulk \cite{kokhanovsky2004a}. Snow is a scattering medium composed of ice grains or crystals, which both scatter and absorb light. The mean free path between scattering events is described by the scattering length \(l_{\mathrm{sca}}=1/\sigma_{\mathrm{sca}}\) or scattering coefficient \(\sigma_{\mathrm{sca}}\). If scattering were absent, absorption is likewise described by the absorption length \(l_{\mathrm{abs}}=1/\sigma_{\mathrm{abs}}\) or absorption coefficient \(\sigma_{\mathrm{abs}}\). Both quantities are properties of the bulk that describe the effective isotropic scattering process with absorption, not the physical process on the level of the individual snow crystal. In the limit of small absorption and an infinitely deep snow pack, the reflectance function \(R(\theta_0,\theta,\phi)\) of the cosines of incident angle \(\theta_0\), observation angle \(\theta\) and relative azimuth \(\phi\) reads \cite{kokhanovsky2004}

\begin{equation}
    R(\theta_0,\theta,\phi) = R_0(\theta_0,\theta,\phi)\mathrm{exp}\left(4\sqrt{\frac{\sigma_{\mathrm{abs}}}{3\sigma_{\mathrm{sca}}}} \frac{K_0(\theta_0)K_0(\theta)}{R_0(\theta_0,\theta,\phi)} \right),
\end{equation}

\noindent where \(R_0\) is the reflectance function of the non-absorbing bulk, and \(K_0\) is the escape function which describes the angular distribution of reflected radiation. To obtain the plane albedo \(\alpha_p\), which describes the amount of radiation reflected for a fixed incident angle, the reflectance function needs to be integrated over all observation and azimuth angles. In snow, the scattering coefficient is typically of the order of \(1...10^3\ \mathrm{m^{-1}}\), while the absorption coefficients of clean ice in the visible spectrum are of the order of \(10^{-3}...\times10^{-1}\ \mathrm{m^{-1}}\). These values correspond to scattering lengths of the order of centimeters, and absorption lengths of the order of several meters. Therefore, \(\sigma_{\mathrm{abs}}/\sigma_{\mathrm{sca}}\) is typically small so that the exponential can be expanded. This yields

\begin{equation}
    \alpha_p(\theta_0) = \frac{2}{\pi}\int_0^{2\pi}d\phi \int_0^1\theta d\theta R(\theta_0,\theta,\phi) \approx 1-K_0(\theta_0)4\sqrt{\frac{\sigma_{\mathrm{abs}}}{3\sigma_{\mathrm{sca}}}}.
    \label{eq:albedo1}
\end{equation}

\noindent The escape function \(K_0\) can be approximated for reasonable acute angles of incidence \cite{sobolev1975}:

\begin{equation}
    K_0(\theta_0) \approx \left( \frac{3}{7} \right) (1+2\theta_0).
\end{equation}

\noindent Further integration over all incident angles yield the spherical albedo \(\alpha\) for diffuse illumination:

\begin{equation}
    \alpha = \frac{2}{\pi}\int_0^1\theta_0 d\theta_0 \alpha_p(\theta_0) = 1-4\sqrt{\frac{\sigma_{\mathrm{abs}}}{3\sigma_{\mathrm{sca}}}}.
    \label{eq:albedo2}
\end{equation}

This is a very convenient description of the albedo and does not depend on any properties of the individual snow grain, only on effective bulk properties. Measurement of these properties makes for a powerful tool for estimating the apparent optical properties of snow one requires for surface energy balance calculations. Recently, these approximations have been found to yield accurate results for snow and can easily be adapted for sloped surfaces \cite{picard2020}.

We have shown in previous work that such measurements are possible in the diffusion far field, where a point source and a detector are separated by many scattering lengths \cite{allgaier2021}. While a completely independent measurement of both parameters has been demonstrated in glacier ice using time-resolve measurement of laser pulses\cite{allgaier2022}, it is challenging for snow because of the shorter scattering length, which makes time-resolving the pulse propagation technically challenging in the field. However, measurements with a stationary light source as proposed in \cite{allgaier2021} can provide a length scale \(L=\sqrt{l_{\mathrm{abs}}l_{\mathrm{sca}}}\). Assuming that the scattering coefficient in the visible spectrum is independent of wavelength, one can use a measurement at two wavelengths to extract the ratio of the absorption coefficients, analog to transmission measurements in snow \cite{warren2008}.

Briefly, the propagation of an instantaneous point-source in an infinite medium is described by the fluence rate \cite{allgaier2021}

\begin{equation}
\Phi(\rho,z,t) = \frac{1}{\left(4\pi Dt\right)^{3/2}}\mathrm{exp} \left(-\frac{\rho^2+z^2}{4Dt}  - c\sigma_{\mathrm{abs}} t\right),
\end{equation}

\noindent with the diffusion constant \( D = c l_{\mathrm{sca}}/3\), speed of light in the medium \(c\), horizontal distance \(\rho\) from the source and vertical distance (depth) \(z\).
Note that it is not necessary to introduce boundary conditions in the case of snow, which is composed of air and ice grains, where the only internal reflections happen inside the grains. This behavior is fundamentally different from internal reflection at a boundary like it can be found in glacier ice, which is typically treated as solid ice with small bubbles responsible for the scattering. We do, however, have to account for the fact that the source is not an isotropic point source, but directional, and replace it with an effective isotropic source located at \(z=-l_{\mathrm{sca}}\). For a constant (stationary source) like a continuous-wave laser focused on the snow, the fluence rate needs to be integrated over all times, yielding the fluence

\begin{equation}
    \Phi(\rho) = \frac{1}{4\pi D} \frac{1}{\sqrt{\rho^2+l_{\mathrm{sca}}^2}}\mathrm{exp}\left(-\sqrt{\frac{3(\rho^2+l_{\mathrm{sca}}^2)}{l_{\mathrm{sca}}l_{\mathrm{abs}}}}\right).
    \label{eq:diffuse}
\end{equation}

\noindent Measuring this distribution on the surface can be done with a camera. The measured data is then fitted with the above function, the exponential part of which yields the constant \(L=\sqrt{l_{\mathrm{abs}}l_{\mathrm{sca}}}\). Even though the function is technically unique to \(D\) and therefore \(l_{\mathrm{sca}}\), the exponential is the dominating part and the scattering length can typically not reliably used as a fit parameter \cite{allgaier2021}.

From there, a differential measurement at two wavelengths \(\lambda,\lambda_{\mathrm{ref}}\) is necessary to determine scattering and absorption coefficients from the two measured quantities \(L_{\lambda},L_{\mathrm{ref}}\). In the field, these length scales typically fall in the range of \(L=0.1...1\) m. We can assume that the size of snow grains, typically in a range of \(0.1...1\) mm, is much larger than the wavelength, implying that scattering is mostly geometric instead of diffractive \cite{kokhanovsky2004}. The only wavelength-dependent parameter in the geometric scattering process is the refractive index of ice, which is almost constant at \(n=1.31\) in the visible spectrum \cite{warren2008}. Therefore, the scattering coefficient is assumed independent of the wavelength, analog to the transmission technique by Warren \cite{warren2006}:

\begin{equation}
    L_{\mathrm{ref}}^2/l_{\mathrm{abs,ref}} = l_{\mathrm{sca,ref}} = l_{\mathrm{sca,\lambda}} = L_{\mathrm{\lambda}}^2/l_{\mathrm{abs,\lambda}}.
    \label{eq:reference}
\end{equation}

\noindent We find for the absorption length at the measurement wavelength \(\lambda\):

\begin{equation}
    l_{\mathrm{abs,\lambda}} = l_{\mathrm{abs,ref}}  \left( \frac{L_{\mathrm{\lambda}}}{L_{\mathrm{ref}}} \right)^2.
    \label{eq:abslength}
\end{equation}

\noindent Unfortunately, the bulk absorption coefficient of snow is not typically known and depends on local and current conditions. What is known is the absorption coefficient of ice for sufficiently large wavelengths, where it is dominated by the absorption of clean ice rather than impurities. The absorption coefficient of snow then depends on properties of both the bulk and the individual grain. Following the derivation by Libois et al. \cite{libois2013} and Kokhanovsky \cite{kokhanovsky2004}, the absorption  coefficient depends on the particle density \(n\) and absorption cross section \(C_{\mathrm{abs}}\) of the individual grain:

\begin{equation}
    \sigma_{\mathrm{abs}} = n C_{\mathrm{abs}} = \frac{d_{\mathrm{snow}}}{d_{\mathrm{ice}}V} C_{\mathrm{abs}},
\end{equation}

\noindent where \(d_{\mathrm{snow}}\) and \(d_{\mathrm{ice}}\) are the density of the snow and ice, respectively, and \(V\) is the volume of the grain. As we will see, the absorption coefficient is independent of the size or volume of the grains, and merely depends on the grain shape and snow density. The absorption cross section reads

\begin{equation}
    C_{\mathrm{abs}} = V \sigma_{\mathrm{abs,ice}} B,
\end{equation}

\noindent where \(\sigma_{\mathrm{abs,ice}}\) is the absorption coefficient of ice, which is typically known for sufficiently large wavelengths, and the absorption enhancement coefficient \(B\). The factor \(B\) describes how much of the optical path is inside the snow grain (where absorption occurs), thus accounting for internal reflections. \(B\) is typically within a range between \(1.25\) and \(2\) and has been calculated for various grain shapes \cite{libois2013}. This range has since been supported by experimental observations \cite{libois2014}, with an average of around 1.4. Therefore, the absorption coefficient of snow is independent of grain size and volume:

\begin{equation}
    \sigma_{\mathrm{abs}} = \frac{d_{\mathrm{snow}}}{d_{\mathrm{ice}}} \sigma_{\mathrm{abs,ice}} B.
    \label{eq:abscoeff}
\end{equation}

\noindent Combining Eq. (\ref{eq:reference}), (\ref{eq:abslength}) and (\ref{eq:abscoeff}) we find that the shape-dependent factor \(B\) as well as the density drop out as they are identical at both wavelength:

\begin{equation}
    l_{\mathrm{sca}} = \frac{L_{\mathrm{ref}}^2 d_{\mathrm{snow}} B}{ l_{\mathrm{abs,ice,ref}} d_{\mathrm{ice}}},
    \label{eq:sca}
\end{equation}
\begin{equation}
    l_{\mathrm{abs,ice,\lambda}} = l_{\mathrm{abs,ice,ref}} \left( \frac{L_{\mathrm{\lambda}}}{L_{\mathrm{ref}}} \right)^2,
    \label{eq:abs}
\end{equation}

\noindent or in terms of the absorption coefficients:

\begin{equation}
    \sigma_{\mathrm{abs,ice,\lambda}} = \sigma_{\mathrm{abs,ice,ref}} \left( \frac{L_{\mathrm{ref}}}{L_{\mathrm{\lambda}}} \right)^2.
    \label{eq:absc}
\end{equation}

\noindent We see that the absorption length at the measurement wavelength depends only on the measured length scales and the absorption length of ice at the reference wavelength, which is known for sufficiently long wavelengths \cite{warren2008}. The scattering length for both wavelengths depends on one measured parameter, density and absorption enhancement factor. While densities are easily measured in the field, observations of snow crystals require practice and good reference photographs \cite{lachapelle1969}. With these parameters known, one can estimate albedo using Eq. (\ref{eq:albedo1})-(\ref{eq:albedo2}) and (\ref{eq:sca})-(\ref{eq:abs}).

In addition to spectral albedo, the absorption coefficient use useful for estimating the concentration of LAPs. As an example, we express the absorption entirely in terms of an effective BC concentration, which implies that BC is the only impurity, an assumption that is certainly untrue. Nevertheless, such a number can be used as an input to physical albedo models such as SNICAR \cite{flanner2021}.  The effective BC concentration \(C\) can then be retrieved from the measured absorption coefficient of ice at the measurement wavelength \cite{doherty2014} and mass absorption coefficient of BC \(C_{BC}\):

\begin{equation}
    C = \frac{\left(\sigma_{\mathrm{abs,ice,\lambda}} - \sigma_{\mathrm{abs,ice,clean}}\right)}{d_{\mathrm{snow}}C_{BC} (\lambda/\lambda_{\mathrm{ref}})^{-1.1}}.
    \label{eq:bc}
\end{equation}

\noindent While it is certainly an unrealistic simplification to assume BC as the only impurity, but doing so will provide us with an upper bound on the actual BC concentration. For a compact overview of the optical properties of snow, see \cite{kokhanovsky2004}.

The method of using the scattering distribution on the surface after injecting a laser beam to infer scattering and absorption coefficients bares some resemblance to oblique incidence reflectometry (OIR), which can be implemented in a similarly compact manner \cite{lin1996} and even using a smartphone camera \cite{cao2021}. However, there are some key differences. OIR uses oblique incidence at angles of \(\approx45^{\circ}\). The scattering distribution is centered around an effective source, similar to Eq. \ref{eq:diffuse}. The oblique incidence allows to establish the distance between the entry point and the effective source, which is directly connected to the mean-free path and therefore the scattering length. However, this is challenging in snow. Due to the rough surface, the entry point is not easily determined with sufficient accuracy. Further, the refractive index would be determined by that of ice, and the density of the snow, \(n_{eff}=n_{air}+((n_{ice}-n_{air})d_{snow}/d_{ice})\), which is highly inhomogeneous at the very top few millimeters of the snowpack due to bunching of snow grains into larger clusters. Here, we use almost normal incidence (incidence angle of \(\approx10^{\circ}\)), which places the effective source close to the entry point. This simplifies image analysis. While this means that the scattering length cannot be determined independently, the particular conditions present in snow and ice, namely the wavelength-independence of the scattering length allow us to circumvent this issue by using two wavelengths (see Eq. (\ref{eq:sca})).

\section{Experimental setup}

\begin{figure}
    \centering
    \includegraphics[width=\textwidth]{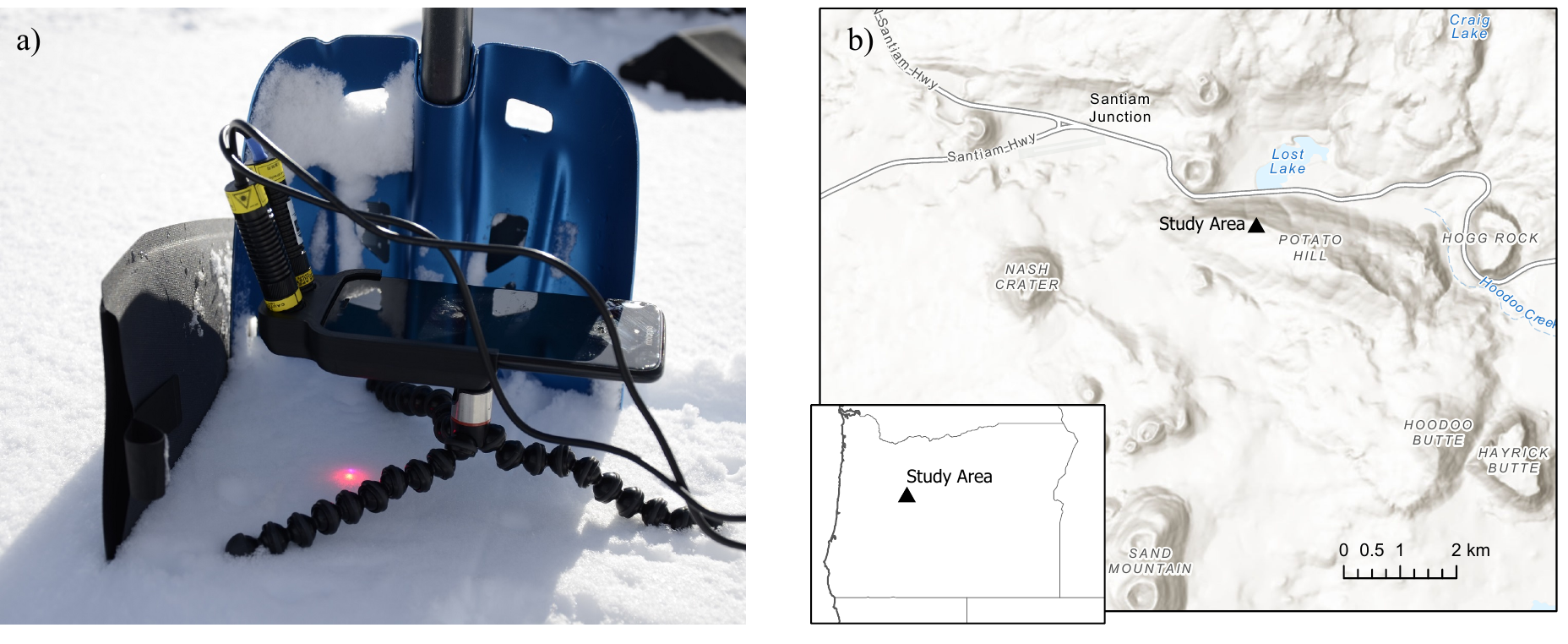}
    \caption{Left: The instrument in the field. The phone is held in a 3D-printed cage, which attaches to a standard camera tripod. The cage also holds the laser diode modules, which are angled so that the spot on the snow is roughly centered below the camera. Right: Location of the study area on Potato Hill, Oregon, USA.}
    \label{fig1}
\end{figure}

The experimental implementation of this measurement requires two spectrally narrowband light sources, in our case two continuous wave diode laser modules emitting at 650 nm (\textit{Laserland 1668-650D-100-5V}) as the reference wavelength and 405 nm (\textit{Laserland 1668-405D-100-5V}) as the measurement wavelength. The measurement wavelength of 405 nm was chosen since it is close to the minimum absorption wavelength of pure ice, where the sensitivity to detect increased absorption due to LAPs is the highest. Both lasers emit approximately 70 mW of continuous power and are powered by a small portable 5V USB battery. Detection is carried out using the camera of a low-end smartphone (\textit{Motorola G7 Play}). Any modern smartphone that gives access to the camera's raw image files (supported on some Android phones running Android 5.0-7.1, most Android phones running Android 8.0 or higher, or Apple iOS 10 or higher) can be used. Raw files allow for reproducible linearity of the imaging sensor's response. The lasers are mounted to the phone using a 3D-printed cage which also acts as a tripod adapter. The setup is shown in Fig. \ref{fig1}b as it is deployed in the field. The two lasers are shown mounted to the top of the phone case on the left of the apparatus. The battery is not shown. On some phones, 5V power can be drawn directly from the phone's USB port. Matlab checkerboard camera calibration patterns are used to obtain a camera calibration and convert pixel coordinates to real-world units. Image exposure is done automatically, centered onto the laser spot on the snow. The imaging sensor will most likely still saturate at the center, but only the outer portion of the image is required for analysis. Snow crystal shape is evaluated with a loupe and snow crystal card with a 1x1mm grid. Snow density is measured with a pocket balance with 0.1 g precision (\textit{Ohaus Gold Series}) and a 0.25 L-volume snow sampler (\textit{Snowmetrics}). Items like an avalanche shovel can be used to shade the study area for better background suppression. 

While it may seem beneficial to use spectral bandpass filters around the laser wavelengths to reject background light, such a filter would be expensive compared to the otherwise low cost of the described components. We find rejection of background light is already sufficient for several reasons. First, the lasers are fairly bright at 70mW average power, which creates significant contrast between the laser light and any background light. Second, we only use the pixels of the RGB filter matrix that match the laser color, effectively rejecting 2/3 of the incident solar background spectrum. Combined, these two effects push the spatially constant background light close to the noise florr of the camera sensor. Lastly, the diffusion model is insensitive to a constant offset. The background light level is both subtractednd left as a free additive fit parameter.

In addition to the differential diffuse imaging experiment, the spectral albedo of snow was recorded using a spectrometer for the visible spectrum with a cosine-corrected receptor attached to the optical fiber (\textit{Ocean Insight USB4000 and CC-3 cosine receptor}). The fiber was attached to a leveled ski pole. 

\section{Data retrieval}

\begin{figure}
    \centering
    \includegraphics[width=0.9\textwidth]{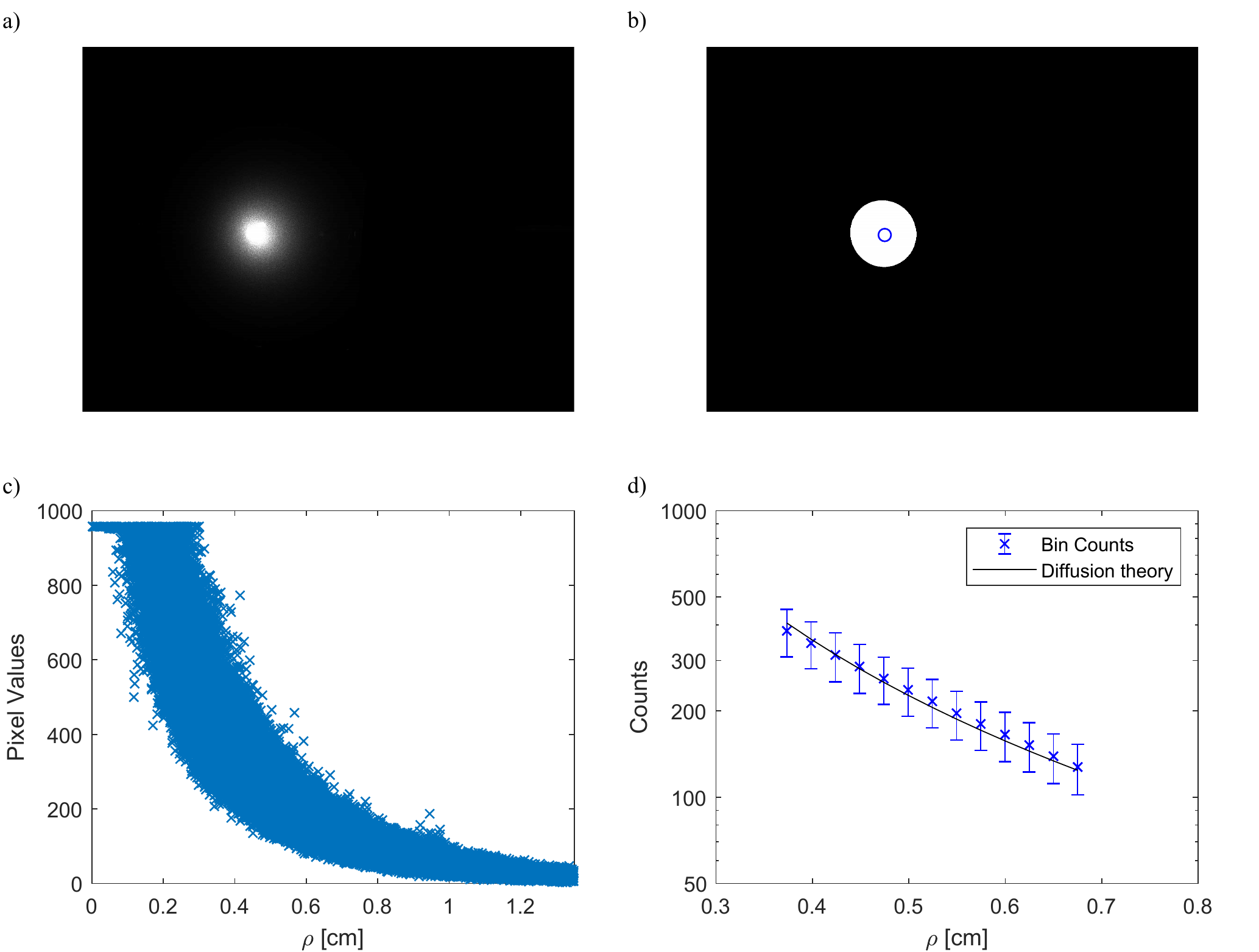}
    \caption{The data retrieval is performed in four steps. a) After conversion into a generic, uncompressed digital negative (DNG) file, only the pixels corresponding to the employed color (red or blue) are used. b) Blurring and applying a threshold transforms the image from panel (a) into a monochrome image. The centroid of this image is found automatically. c) All pixels are resorted according to their distance from the center. A camera calibration derived from a separate checkerboard calibration image is used to convert distances to real-world units. d) The pixels are averaged in predefined bins (here 0.25 mm wide). The used range extends from the smallest distance at which clipping occurs to the largest distance at which a pixel reads zero counts. The averaged values are used to fit the diffusion theory (Eq. (\ref{eq:diffuse})), which yields the scaling constant \(L\).}
    \label{fig2}
\end{figure}

The captured raw image files are converted into a generic, uncompressed digital negative format (DNG) using the free \textit{Adobe DNG Converter}.
Data analysis is performed in Matlab, using components from the Image Processing Toolbox. The code is supplied as supplementary material. Fig. \ref{fig2} illustrates the data analysis procedure. Fig. \ref{fig2}a shows all red pixels of an acquisition using the red laser and a scattering sample in the laboratory. The other pixels from the Bayer matrix are discarded. Next, the center of the image is detected. For this, we apply a Gaussian filter and convert it into a monochrome image and as shown Fig. \ref{fig2}b. From this, the centroid can be extracted. Using a calibration obtained from a separate checkerboard calibration card, all pixels are assigned coordinates in real world units and resorted into a linear array of ascending distance to the center as shown in Fig. \ref{fig2}c, from which background is subtracted. Identifying the farthest saturated pixel and the closest pixel with value zero, we establish a range of data that will be used for fitting with Eq. (\ref{eq:diffuse}). To do this, the data is averaged over 0.25 mm-width intervals. the scaling constant \(L=\sqrt{l_{\mathrm{sca}}l_{\mathrm{abs}}}\) is obtained from the fit. The fit suffers from a fairly flat parameter landscape, making the individual extraction of scattering and absorption length ambiguous, but in principle the fit retrieves \(L\) reliably from simulated data \cite{allgaier2021}. While setting initial parameters for the fit is crucial, each data set required only one set of these initial parameters, and fit results are only used if significant deviation of the retrieved fit parameters from initial parameters is observed. Fit convergence is verified by calculating the normalized square error of the fit. Due to surface roughness, noise, or hot pixels, establishing an adequate range of data to pass to the fit is sometimes unsuccessful. Under field conditions, it is common for the checkerboard detection to fail due to uneven lighting or specular reflection obscuring the checkerboard corners. These issues will trigger the rejection of the data point.

\section{Calibration measurements using scattering phantoms}

\begin{figure}
    \centering
    \includegraphics[width=\textwidth]{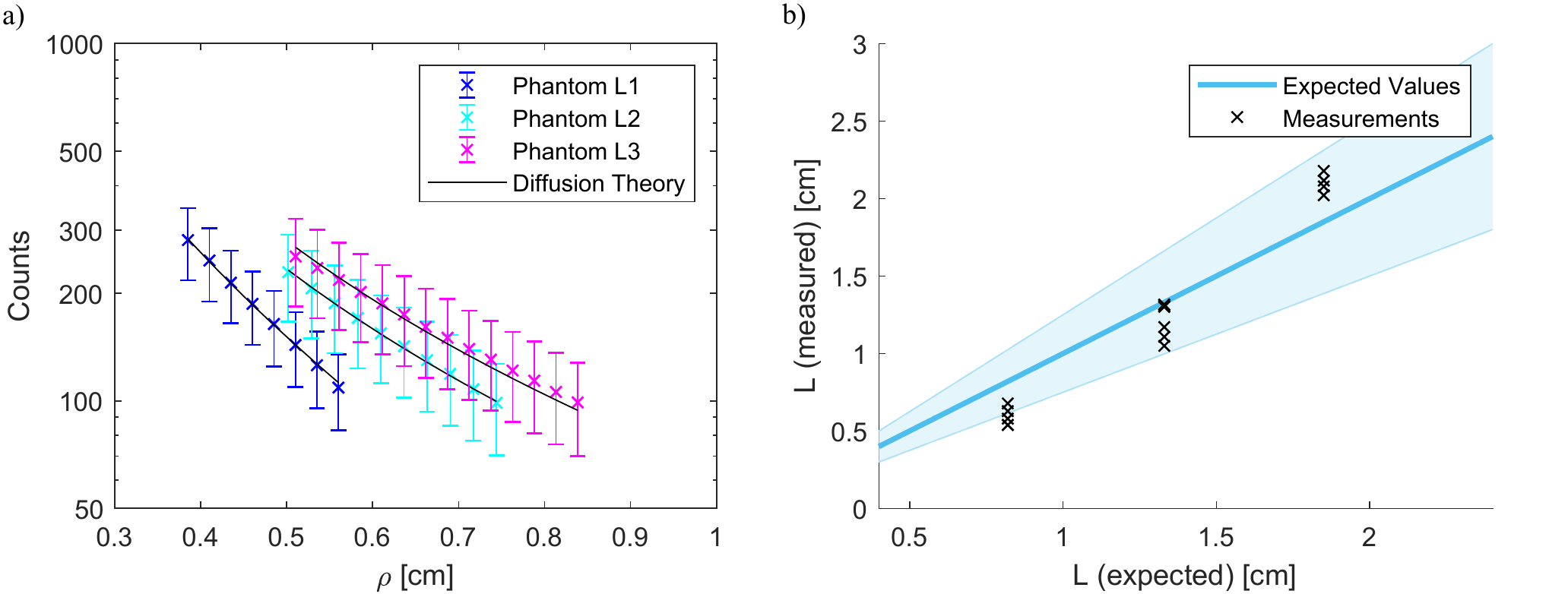}
    \caption{Measurement of the fall-off constant for three known reference phantoms using 8 measurements each. a) Examples for count fall-off for one measurement of each of the three reference phantoms L1, L2 and L3 with corresponding curve fits. b) Extracted fall-off constants from all 24 measurements. The blue line follows \(L_{\mathrm{measured}}=L_{\mathrm{expected}}\), the shaded area corresponds to the uncertainties of the nominal attenuation constants. The measured values are placed on the x-axis according to the nominal values L1, L2 an L3 of the manufactured scattering phantoms.}
    \label{fig3}
\end{figure}

For validation of the measurement of the attenuation constant \(L\) we employ scattering phantoms of different known scattering and absorption coefficients in the laboratory. The scattering phantoms are fabricated out of (\textit{Platsil SiliGlass}) compound using glass microspheres as scatterers. We follow the recipe suggested by Sekar et al.\cite{sekar2019}, with the exception of the Dye, which was not available to use. We instead used a generic black resin dye and fabricated several samples of different thickness and containing different concentrations of dye but no scatterers. The absorption coefficient was measured for each dye concentration. From these meausurements, we obtained a mixing equation that relates the dye concentration of \(C\) in percent by volume and the absorption coefficient  as \(\sigma_{\mathrm{abs}} = (2.0\cdot C+0.008)\ \mathrm{cm}^{-1}\). Given the range of known parameters of this scattering phantom in terms of scatterer concentration and spectral properties, we can employ the recipe to fabricate samples imitating fine-grained snow at the red end of the spectrum. The results for three scattering phantoms with \(L_1=0.82(\pm0.28)\ \mathrm{cm}\,\ L_2=1.33(\pm0.46)\ \mathrm{cm}\) and \(L_3=1.85(\pm0.65)\ \mathrm{cm}\) are shown in Fig. \ref{fig3}. Eight measurement using the red 650 nm-laser were carried out for each sample. The resulting data and best fit are shown in Fig. \ref{fig3}a, showing the three different slopes. All of 24 measurements are shown in Fig. \ref{fig3}b and fall into the range in which the attenuation constant \(L\) is known. The mean retrieved scaling constants for each sample are \(L_1=0.82(\pm0.28)\ \mathrm{cm}\,\ L_2=1.33(\pm0.46)\ \mathrm{cm}\) and \(L_3=1.85(\pm0.65)\ \mathrm{cm}\).

To imitate snow at the blue end of the visible spectrum is not possible, because we lack a calibration of the recipe's scattering coefficient in the blue. In addition, such a sample would require minuscule concentrations of the scattering spheres, respectively, as well as a large volume of the order of 1 \(\mathrm{m^3}\), which is not feasible or accurate. While the scattering lengths in the used scattering samples are 0.2 cm, 0.35 cm and 0.5 cm, respectively - comparable to that measured in snow - the material absorption coefficient is much larger than that of real snow. We will see that the values for \(L\) encountered on snow are of the order of \(0.1...10\) m due to the small scattering coefficient of snow and ice. Therefore, the values for \(L\) in these scattering phantoms is not representative of fresh, low-density snow and about 1 order of magnitude too small. However, the magnitude of the scattering constant of the samples is the same as that found in snow. Therefore, the test involving scattering phantoms serves as a first indication that the attenuation constant is reliably retrieved using the technique.
For field measurements, an appropriate setting with well-defined parameters (like what we find in fresh snow) can help with further validation, as the scattering and absorption coefficients as well as albedo can be estimated in some cases.

\section{Field measurements on the Oregon Cascade crest}

Field work was conducted on December 17th 2021 on Potato Hill (44.425519\(^{\circ}\)N,121.911703\(^{\circ}\)W, Elevation 1443m - see Fig. \ref{fig1}b) in the Cascade Range in Oregon, USA. The study area is facing South-West with a slope angle of 10\(^{\circ}\) and is located in a forest clearing from a recent wildfire, located close to the main highway crossing the Cascade Range in Oregon. During spectral albedo measurements, the sky was covered by clouds, providing diffuse illumination. Snow had fallen two days prior to field work, and was found to contain slightly rimed, star-like snow crystals 1mm in diameter as well as smaller fragments. We assume a shape-dependent absorption-enhancement coefficient of \(B\)=1.84. This value has been calculated for star-like (fractal) snow crystals \cite{libois2013} but is also fairly accurate for small spheres \cite{kokhanovsky2004}, which can therefore be used as a reasonable approximation. The snow density was measured at 125.9 \(\mathrm{kg}\cdot\mathrm{m}^{-3}\). For the reference wavelength of 650 nm, we use a value of \(\sigma_{\mathrm{abs,ice,ref}}=2.6\times10^{-1}\ \mathrm{m^{-1}}\) that we obtained using an independent measurement on Collier Glacier (also in the Central Oregon Cascades and the closest reference point to our knowledge) \cite{allgaier2022}.

Using the smartphone sensor, a series of 23 measurements were performed. We excluded those where either the camera calibration failed (calibration algorithm failed to detect the checkerboard corners), the algorithm was unable to establish a range of usable (unsaturated and noise-free) data, or the least-square fit did not converge, which leaves us with 10 measurements. The successful measurements yield a mean scattering coefficient of \(2.13(\pm0.16)\times10^{2}\ \mathrm{m^{-1}}\) (using Eq. (\ref{eq:sca})). The corresponding scattering length is \(4.7(\pm0.4)\) mm. Accounting for the strongly forward peaked scattering of snow crystals with a typical aniostropy factor \(g=0.89\) \cite{kokhanovsky2004}, the corresponding mean-free-path is \(l_{\mathrm{sca}}(1-g)=0.52(\pm0.5)\) mm, close to the crystal size estimated by eye.

\begin{figure}
    \centering
    \includegraphics[width=\textwidth]{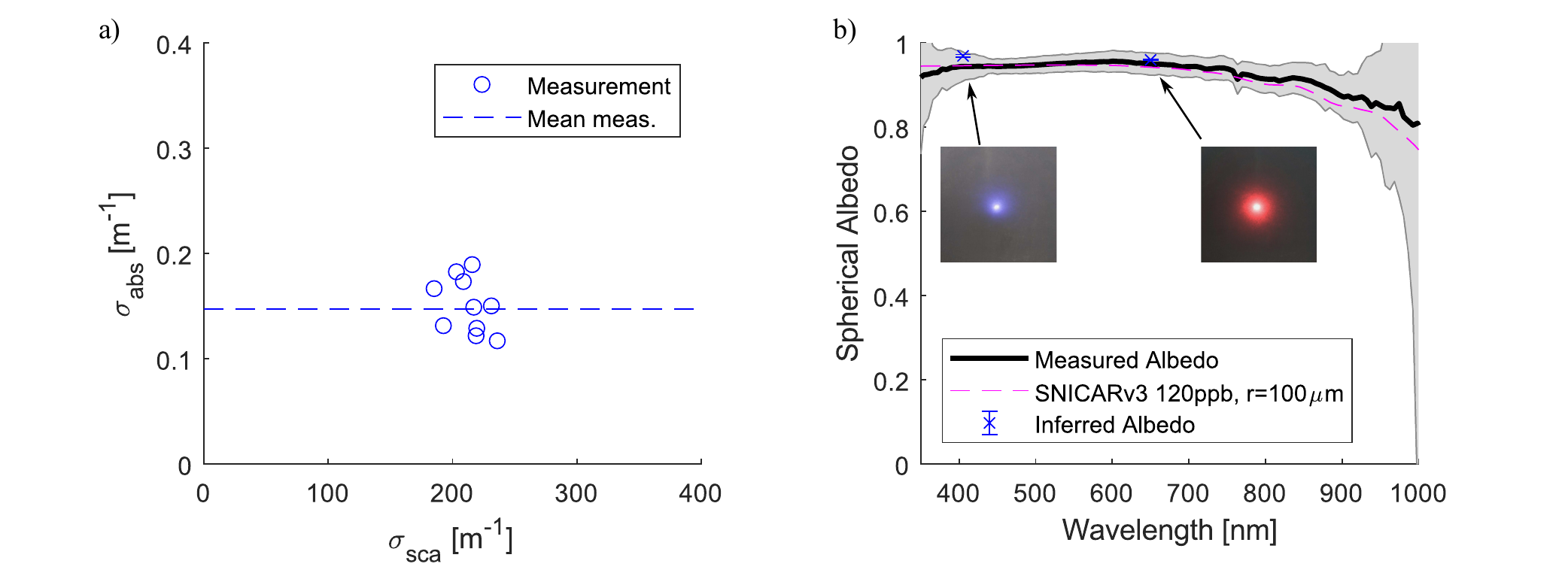}
    \caption{Results from field measurement. (a) Scattering and absorption coefficient from each successful measurement. The blue dashed line represents the mean absorption coefficient at 405nm from the measured data. (b) Spherical spectral albedo from direct measurement in the field (black line), simulated using SNICARv3 and a BC concentration of 120pbb (magenta), and calculated using the measured scattering and absorption coefficients. Insets show a measurement photograph at each of the two wavelength.}
    \label{fig4}
\end{figure}

In addition to the scattering coefficients, we obtain the absorption coefficient of ice for each measurement as shown in Fig. \ref{fig4}a using Eq. (\ref{eq:absc}). The average value is \(\sigma_{\mathrm{abs,ice,\lambda}} = 1.5(\pm0.3)\times10^{-1}\ \mathrm{m^{-1}}\). This indicates that there is significant contamination with LAPs present at the field site. Using Eq. (\ref{eq:bc}), we obtain an upper bound on the BC concentration of 120(\(\pm\)20) pbb. While the true BC concentration is certainly lower, the number provides some context and can act as an input to SNICAR for simulating the albedo at the measurement wavelength. As long as only this wavelength is of interest, the model is agnostic to the type of LAP that is provided as an input, given that it corresponds to the correct increase of the absorption coefficient.

In addition to scattering and absorption coefficients, it is now possible to calculate plane and spherical albedo for both the reference wavelength (using the measured scattering coefficient and reference ice absorption coefficient) and the measurement wavelength (using both measured scattering and ice absorption coefficient) using Eq. (\ref{eq:albedo1})-(\ref{eq:albedo2}). The resulting values for the spherical albedo are shown in Fig. \ref{fig4}b along with the spherical (diffuse or white-sky) albedo measured in the field and simulated albedo using SNICAR-AD v3 \cite{flanner2021} (assuming the concentration of 120 ppb as extracted from the measured absorption coefficient, and an equivalent spherical particle size of 100\(\mu\) m estimated to match the scattering cross-section of larger, star-like snow crystals). The grey shaded area represents the standard deviation of the albedo measurement obtained from 10 acquisitions. At both 405 and 650 nm, the inferred albedo is comparable to the directly measured spectral albedo. The simulated albedo using the inferred BC concentration agrees with the measured spectral albedo.

While this can serve as a first demonstration of this measuremment technique and its ability to provide estimates of LAP concentration and spectral albedo, further validation is needed to evaluate instrument performance over a larger range of LAP concentrations. Laboratory analysis of snow samples would have the benefit of providing the correct ratio of actual LAPs (BC vs. dust vs. algae).

For dealing with an inhomogeneous medium such as snow, diffusion is well suited to retrieve average properties of the volume. Here, the scattering coefficient of snow is \( 2\times10^2\ \mathrm{m^{-1}}\), those of the scattering phantoms lie in the range of \(2...5\ \times10^2\ \mathrm{m^{-1}}\). In our previous work, we investigated the applicability of the measurement numerically using typical parameters of glacier ice, with scattering coefficients of the order of \(0.6...2\ \mathrm{m^{-1}}\). In those simulations, parameter retrieval was reasonable resilient against the presence of a boundary of varying absorption \cite{allgaier2021}. For time-resolved diffusion measurements, retrieval was successful for scattering coefficients in similar range of \(0.1...5\times10^1\ \mathrm{m^{-1}}\) under rel world conditions. In the case of OIR, others were successful in using a similar model to retrieve scattering coefficients of the order of \(2...8\times10^2\ \mathrm{m^{-1}}\) in different media such as apple, milk and tissue \cite{cao2021}, implying that diffusion models such as the one used here are reliable over a large range of parameters and independent of the actual medium and nature of the scattering process. The same is true for the absorption coefficient, which can range from \(1\times10^{-3}\ \mathrm{m^{-1}}\) in clean ice to \(5\times10^1\ \mathrm{m^{-1}}\) in tissue. Using this model for the measurements presented here does not require absolute intensity measurement, making it impervious to spatially constant background.

\section{Discussion}

So far, field measurements of the intrinsic optical properties of snow are scarce. There are those conducted in the Arctic regions \cite{warren2006}, but little work exists outside of the Arctic, where snow plays a major role in the water cycle. While measuring the apparent optical properties is somewhat more straightforward using pyranometers or spectrometers, these measurements typically only consider few sites or only one at a permanent installation. Even then, extracting the intrinsic optical properties of snow, namely the bulk scattering and absorption coefficients, is hardly possible from reflectance spectroscopy alone, given the large ambiguity of the spectral albedo in respect to grain shape and size, scattering phase function, density and particulate concentrations. The measurement technique presented here can not only provide a direct measurement of the scattering and absorption coefficient at any visible wavelength, but can at the same time provide estimates for the albedo for any lighting condition - plane or spherical radiation or an arbitrary mix of the two for any zenith angle can be calculated. Particularly the absorption coefficient provides a convenient proxy for estimating effective particulate concentrations in the field, without the need for chemical sample analysis. The technique could potentially be used to infer water content and mixing conditions recently obtained using reflectance spectroscopy \cite{donahue2022} with little required modeling.

The magnitude of the snow properties measured here - scattering coefficient and absorption coefficient - is consistent with the expected range and visual observations. The scattering coefficient of \(213(\pm16)\ \mathrm{m^{-1}}\) corresponds to a scattering-mean-free path of \(l_{\mathrm{sca}}/(1-g)=0.52(\pm0.5)\) mm, which is consistent with the millimeter-sized crystals observed by eye. Similarly, the extracted ice absorption coefficient at 405 nm of \(\sigma_{\mathrm{abs,ice,\lambda}} = 1.5(\pm0.3)\times10^{-1}\ \mathrm{m^{-1}}\), falls in the range observed on nearby Collier Glacier \cite{allgaier2022}. The corresponding effective concentration of BC of 120(\(\pm\)20) pbb is consistent with ice showing typical concentrations of black carbon found in North America \cite{doherty2014}.

The core strength of this method, however, lies in its simplicity, portability and cost. Adding the price for the phone (\$120), the laser diode modules (\$20 each), scale (\$70), snow sampler (\$100) and loupe (\$10), the cost for the entire setup is only \$340. Using a household scale and any improvised snow sampler of known volume (piece of PVC pipe, soup can) and assuming one already owns the phone, the method can be implemented for \$50. Therefore, this method is viable for undergraduate research, high school education, and citizen science. The entire setup is compact and lightweight and can therefore easily be transported on foot for deployment in rugged and mountainous terrain.

The combination of the variety in provided information, low cost and portability makes this method an excellent tool for large-scale field campaigns and education alike.

\section{Backmatter}

\begin{backmatter}
\bmsection{Funding}
This work was funded by the University of Oregon through the \textit{Ren\'{e}e James seed grant initiative}

\bmsection{Acknowledgments}
We thank Susan Kaspari for fruitful discussions.

\bmsection{Disclosures}
The authors declare no conflicts of interest.

\bmsection{Data availability} All raw images, reflectance spectra, derived data and the matlab code for image analysis and albedo calculation are published at https://doi.org/10.5281/zenodo.6095959. An online version of SNICAR-ADv3 \cite{flanner2021} can be found at http://snow.engin.umich.edu/.

\end{backmatter}

\bibliography{snow}

\end{document}